\documentclass{appolb}
\usepackage{epsfig}
\usepackage{amssymb}
\usepackage{amsmath}
\usepackage{subfigure}

\newcommand{\fmn}[2]{\mbox{${\textstyle \frac{#1}{#2}}$}}

\begin{document}
\date{\today}
\pagestyle{plain}
\newcount\eLiNe\eLiNe=\inputlineno\advance\eLiNe by -1
\title{$\eta$-MESON PRODUCTION IN NUCLEON-NUCLEON COLLISIONS%
\thanks{Presented at the International Symposium on Mesic Nuclei, Krak\'ow, June 16, 2010}%
}
\author{Colin Wilkin
\address{Physics and Astronomy Dept., UCL, Gower Street, London, WC1E
6BT, UK }} \maketitle

\begin{abstract}
New experimental data are presented on the energy dependence of the
total cross sections for the $np\to d\eta$ and quasi-two-body $pp\to
pp\eta$ reactions, where the final diproton is detected at very low
excitation energy. Differential cross sections of $pp\to pp\eta$ away
from threshold show the influence of higher partial waves and a
partial wave decomposition is attempted.
\end{abstract}

\section{Introduction}

If one looks in a dictionary one sees that the original definition of
a symposium was ``A drinking party at which there was intellectual
conversation''. Further research reveals that, in a game sometimes
played at symposia, participants swirled the dregs in their wine
glasses and flung them at a target --- but not, apparently, in
Krak\'ow!

The possibility that the $\eta$ meson might form some kind of
quasi-bound state with a nucleus has excited physicists for many
years. It was first suggested by Haider and Liu~\cite{HAI1986}, on
the basis of the $\eta$-nucleon information available in 1986, that
the lightest nucleus where this was likely to happen was $^{12}$C.
However, no unambiguous signals for the production of such a state
below the $\eta A$ threshold have ever been found, in part due to the
enormous hadronic backgrounds associated with non-$\eta$ events.

On the other hand, if the $\eta A$ pole is close to threshold it
should influence the $\eta A$ production data a little above
threshold, in much the same way that the deuteron bound state
dominates $S$-wave neutron-proton spin-triplet scattering. It was
therefore argued~\cite{WIL1993} that the available data on
$pd\to\,^{3}$He$\,\eta$ might signal the formation of
$_{\eta}^{3}$He. The obvious drawback of such an approach is that low
energy production data can never determine whether there is a bound
state (e.g.\ the deuteron) or a virtual state (e.g.\ the spin-singlet
$NN$ system). One possible way of resolving this ambiguity is to
study the $A$-dependence of low energy $\eta$ production and use the
fact that the binding should increase with atomic
number~\cite{WIL1997}. I shall here show new results in the $A=2$
systems and leave others to discuss $A>2$.

\section{The $\boldsymbol{\eta d}$ interaction}

The $\eta d$ interaction was studied at CELSIUS in two different
kinematical regimes, \textit{viz} quasi-free $pd \to p_{\rm sp}
d\eta$~\cite{CAL1997} and the same reaction at much lower energies,
\textit{i.e.}\ well below the threshold in nucleon-nucleon
collisions~\cite{BIL2002}. A consistent \textit{FSI} description of
both data sets was achieved by dividing the $\eta d$ invariant mass
distribution by the corresponding phase space.

\begin{figure}[htb]
\begin{center}
\includegraphics[clip,width=6cm]{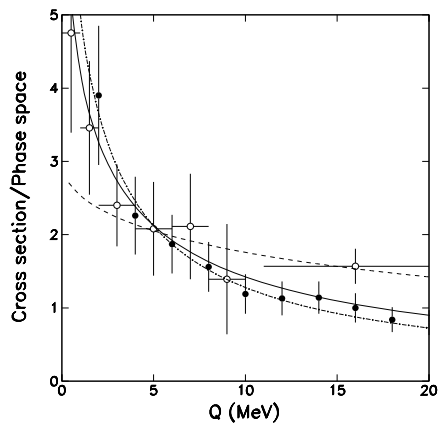}
\caption{Ratio of the cross section for the production of the
$d\,\eta$ system to phase space, as a function of the excitation
energy in the $d\,\eta$ rest frame, for the $pn\to d\,\eta$ total
cross section (open circles)~\cite{CAL1997} and the $pd\to pd\,\eta$
reaction at 1032~MeV (closed circles)~\cite{BIL2002}. The broken,
solid and chain curves are the predictions of the scattering length
formula of Eq.~(\ref{FSI}) using $a_{\eta d}=(0.73+0.56i)$~fm,
$(1.64+2.99i)$~fm, and $(-4.69+1.59i)$~fm,
respectively~\cite{SCH1998}. The overall normalisations are
arbitrary. \label{etadfsi}}
\end{center}
\end{figure}

The results obtained from the two CELSIUS experiments are shown in
Fig.~\ref{etadfsi} in terms of the $\eta d$ excitation energy, where
it is seen that the ratios drop by about a factor of four over
10~MeV. This is an indication of a strong final state interaction
(\textit{FSI}) which, in scattering length approximation, may be
written as
\begin{equation}
\label{FSI} \textit{FSI} \propto 1/\left|1-ika_{\eta d}\right|^2\,
\end{equation}
where $k$ is the $\eta d$ relative momentum.

Three-body estimations of the complex scattering length $a_{\eta d}$
depend critically on the $\eta$-nucleon input~\cite{SCH1998} and
calculations made using modest, strong, and very strong $a_{\eta N}$
values are shown in Fig.~\ref{etadfsi}. The very strong case leads to
a quasi-bound $\eta d$ state whereas in the strong case it is merely
a virtual state. Above-threshold data cannot distinguish between
them.

\begin{figure}[htb]
\begin{center}
\includegraphics[clip,angle=0,width=6.2cm]{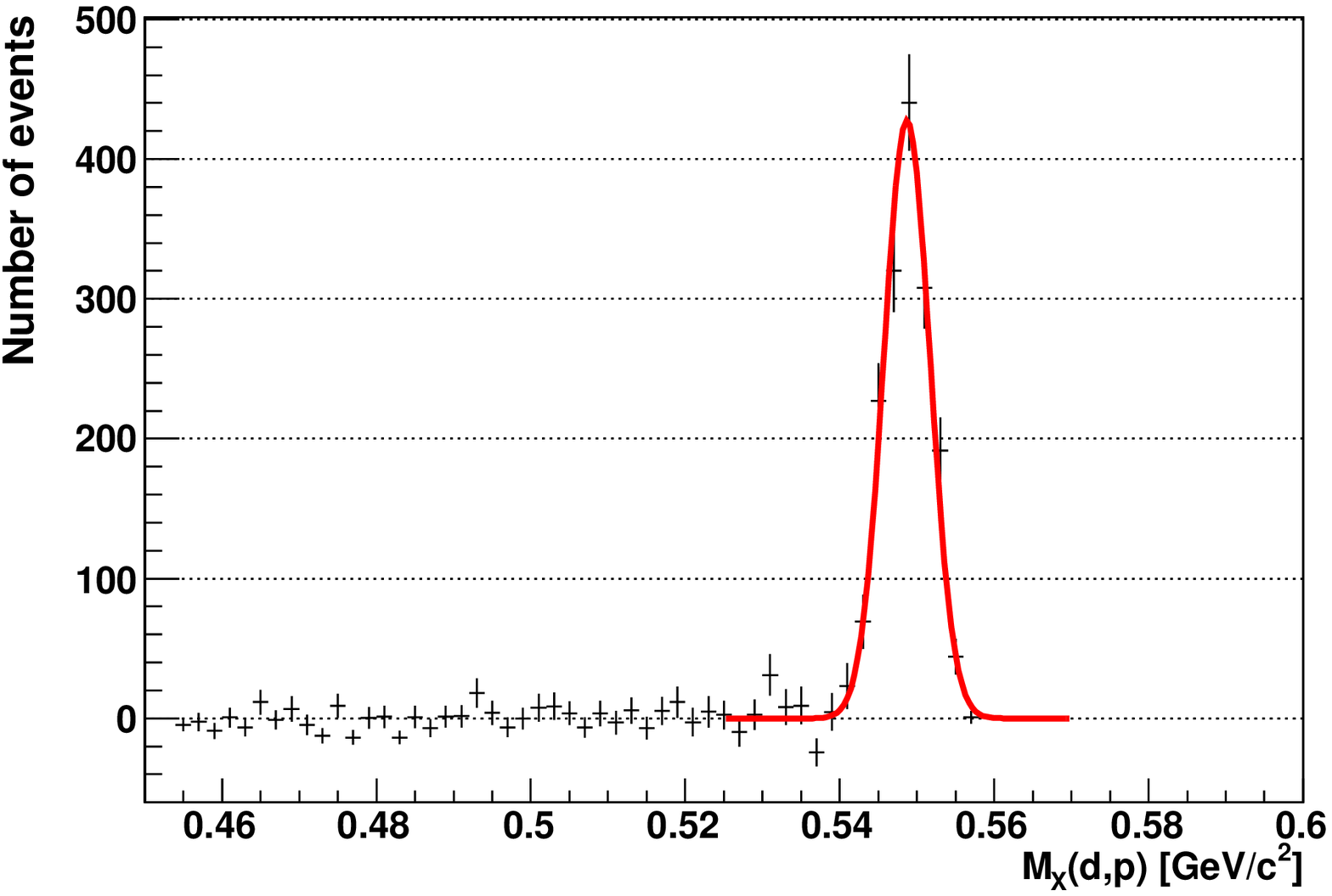}
\includegraphics[clip,angle=0,width=6.2cm]{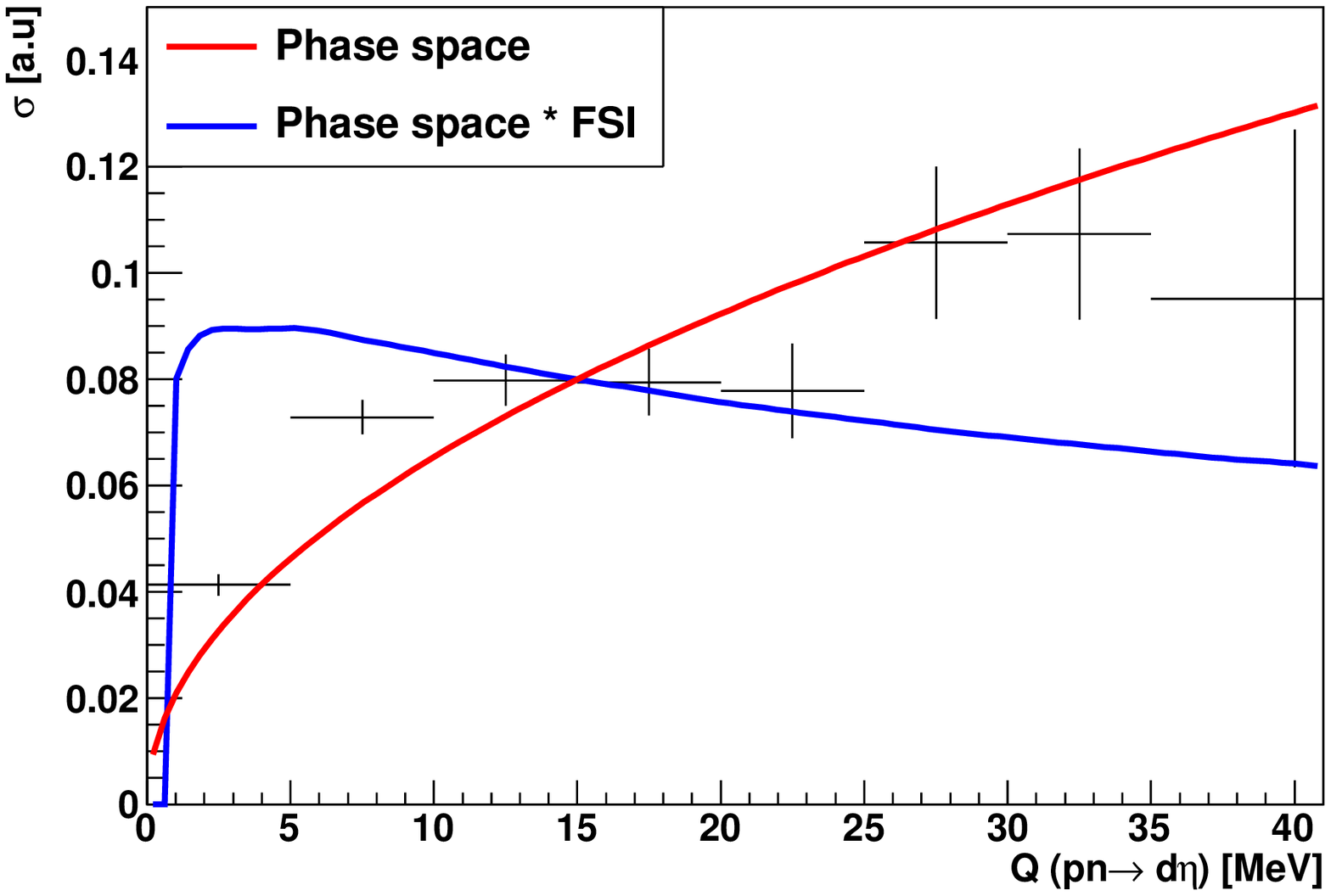}
\caption{Left: Missing-mass spectrum after background subtraction for
the $dp\to d p X$ reaction at 2.27~GeV from preliminary ANKE data for
$5<Q<10$~MeV~\cite{DYM2010}. Right: Preliminary unnormalised $np\to
d\eta$ total cross sections extracted from $dp\to d p X$ data at
2.27~GeV assuming quasi-free production~\cite{DYM2010}. The
phase-space dependence and this modified by the \textit{FSI} with
$a_{\eta d}=(1.64+2.99i)$~fm are shown.\label{pndeta}}
\end{center}
\end{figure}

New data are currently being analysed at COSY-ANKE~\cite{DYM2010}.
When using a deuteron beam, a spectator proton $p_{\rm sp}$ will be
fast and therefore measured in the ANKE forward detector along with
other charged particles. Data were taken on $dp\to p_{\rm sp}dX$ at
the maximum COSY deuteron beam energy of 2.27~GeV and the meson $X$
identified using the missing-mass technique. Since the central
neutron energy is below the $d\eta$ threshold, only the upper part of
the Fermi momentum will contribute to $\eta$ production. The
below-threshold data provide a robust method for background
subtraction and, when this is carried out, the $\eta$ signal shown in
Fig.~\ref{pndeta} is very clean and the whole $\eta$ angular domain
is sampled.

If the data are interpreted in terms of quasifree production, the
energy dependence of the total cross section can be extracted and the
results of this are presented in arbitrary units in
Fig.~\ref{pndeta}. When normalised at 15~MeV, the curve representing
the variation expected on the basis of phase space underestimates the
near-threshold rise but the introduction of the \textit{FSI} with
$a_{\eta d}=(1.64+2.99i)$~fm overdoes things in this region. The
analysis is very preliminary and it is too soon to draw firm
conclusions. However, it is possible that the problem resides in the
quasi-free assumption because, at only a slightly lower energy, it
was claimed that quasi-free production might account for less than
half of the observed cross section~\cite{BIL2002}.

%
%

\section{The $\boldsymbol{pp\to pp\eta}$ reaction away from
threshold}

In order to study the effects of $S$-wave rescattering of the $\eta$
meson from a proton pair, and hence investigate the $\eta pp$
\textit{FSI}, it is important to know at what point higher partial
waves are needed for the description of the $pp\to pp\eta$ reaction.
This means that one has to measure differential observables away from
threshold. The production of the $\eta$ in proton-proton collisions
was investigated by the CELSIUS-WASA collaboration at $Q = 40$ and
72~MeV by detecting the $\eta$ through its $3\pi^0$
decay~\cite{PAU2006}. However, the data from the two-photon decay
branch have been subjected to a much more refined analysis, which is
now approaching completion~\cite{PET2009}.

\begin{figure}[htb]
\begin{center}
\includegraphics[clip,width=6.2cm]{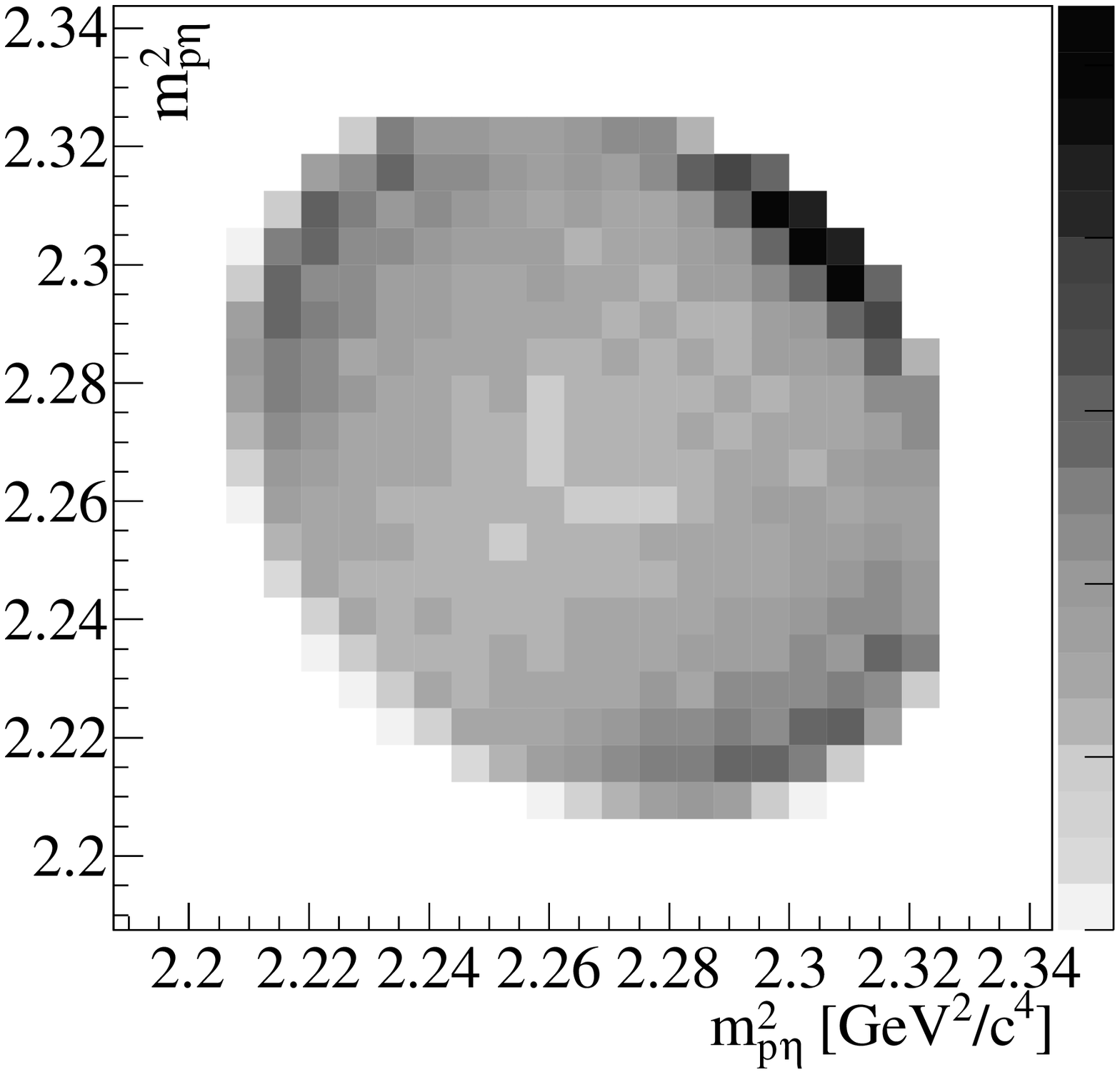}
\includegraphics[clip,width=6.2cm]{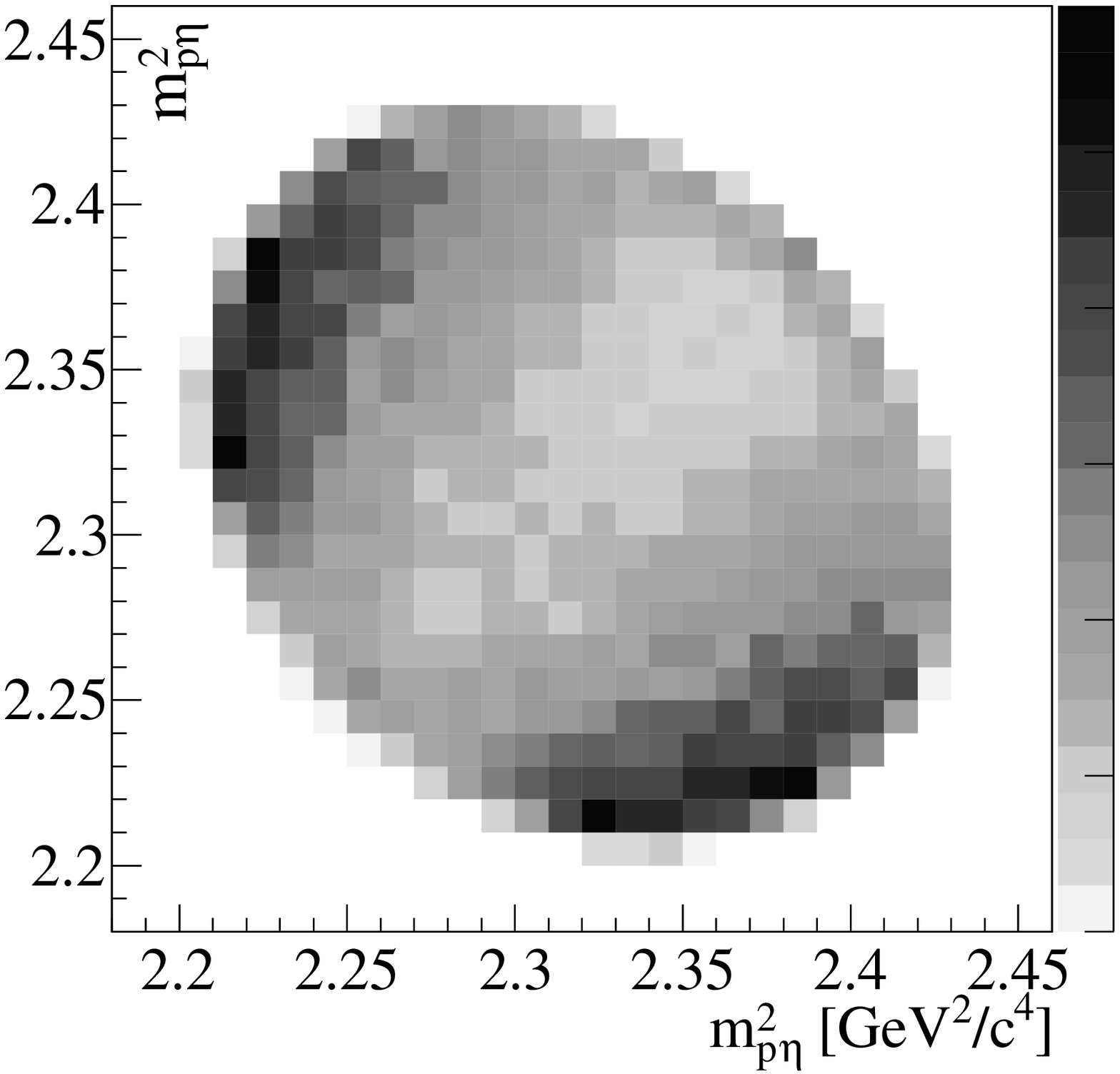}
\caption{Dalitz plots for the $pp\to pp\eta$ reaction at (left)
$Q=40$~MeV and (right) $Q=72$~MeV~\cite{PET2009}. \label{Dalitz}}
\end{center}
\end{figure}

The Dalitz plots for the $pp\to pp\eta$ reaction shown in
Fig.~\ref{Dalitz} look similar when the meson is detected through its
$3\pi^0$~\cite{PAU2006} or $2\gamma$ decay~\cite{PET2009}. The
distributions are far from uniform and the data at both 40 and 72~MeV
show a deep valley along the diagonal where $m(\eta p_1) \approx
m(\eta p_2)$. This is probably due to the $\eta$ being able to form
the $N^*(1535)$ with only one nucleon at a time. The valley implies
that there must be higher partial waves in both the $pp$ and
$\eta{pp}$ systems, at least $L_{pp}\ell_{\eta}=Pp$. It should also
be noticed that the $pp$ \textit{FSI} is only significant at 40 MeV.

Since only the start of the $N^*(1535)$ is sampled, even at 72~MeV,
it is tempting to parametrise the data with partial wave amplitudes
with constant coefficients. Although the $N^*(1535)$ will not be
explicitly included, factors arising from the angular momentum
barriers and the $pp$ \textit{FSI} will.

The square of the $pp\to pp\eta$ matrix element, averaged over
polarisations, is~\cite{PET2009}
\begin{eqnarray}
\label{msq}%
\nonumber \langle |\mathcal{M}|^2\rangle &=& |A_{Ss}|^2
+\fmn{1}{9}|A_{Sd}|^2k^2\left[3(\hat{p}\cdot\vec{k}\,)^2+k^2\right]
+\fmn{1}{9}|A_{Ds}|^2q^2\left[3(\hat{p}\cdot\vec{q}\,)^2+q^2\right]\\
&&\hspace{-2cm}\nonumber
+\fmn{2}{3}\textrm{Re}\left\{A_{Ss}^*A_{Sd}\right\}\left[3(\hat{p}\cdot\vec{k}\,)^2-k^2\right]
+\fmn{2}{3}\textrm{Re}\left\{A_{Ss}^*A_{Ds}\right\}\left[3(\hat{p}\cdot\vec{q}\,)^2-q^2\right]\\
&&\hspace{-2cm}\nonumber
+\fmn{2}{9}\textrm{Re}\left\{A_{Sd}^*A_{Ds}\right\}
\left[9(\hat{p}\cdot\vec{k}\,)(\hat{p}\cdot\vec{q}\,)(\vec{k}\cdot\vec{q}\,)
-3\left(q^2(\hat{p}\cdot\vec{k}\,)^2+k^2(\hat{p}\cdot\vec{q}\,)\right)^2
+k^2q^2 \right]\\
&&\hspace{-2cm} +|A_{Ps}|^2q^2 +2|A_{Pp}|^2(\vec{k}\cdot\vec{q}\,)^2,
\end{eqnarray}
where terms up to fourth order in combinations of the $\eta$
($\vec{k}\,$) and $pp$ relative momentum ($\vec{q}\,$) have been
retained for partial waves up to $D$ in the incident $pp$ system,
where the c.m.\ momentum is $\vec{p}$. The five partial wave
amplitudes $A_{L_{pp}\ell_{\eta}}$ are in the standard notation,
where $L_{pp}$ is the angular momentum in the final $pp$ system and
$\ell_{\eta}$ that of the $\eta$ with respect to the recoiling $pp$.

The momentum factors associated with particular partial wave
combinations are, of course, necessary but the critical assumption in
the analysis is that the $A_{L_{pp}\ell_{\eta}}$ themselves are
constant, apart from the $pp$ \textit{FSI} in the final $S$-wave.
This effect has been taken into account by multiplying the amplitudes
by the ratio of the Paris $pp$ wave function, including the Coulomb
distortion, to the plane wave evaluated at a $pp$ separation of
1~fm~\cite{LAC1980}.

A large variety of one-dimensional data were fitted simultaneously at
40 and 72~MeV. Since the phases of $A_{Ps}$ and $A_{Pp}$ are
irrelevant at this level, this involved the determination of seven
real parameters. The distributions fitted include those in the
invariant masses, the $\eta$ c.m.\ angle, as well as the
Gottfried-Jackson angle. The fits were made to the raw spectra, after
the model had been passed through the full simulation of the
CELSIUS-WASA set-up. However, the results shown here are in terms of
cross sections, where the acceptance has been evaluated using the
model with the best-fit parameters.

\begin{figure}[htb]
\begin{center}
\includegraphics[clip,width=6.2cm]{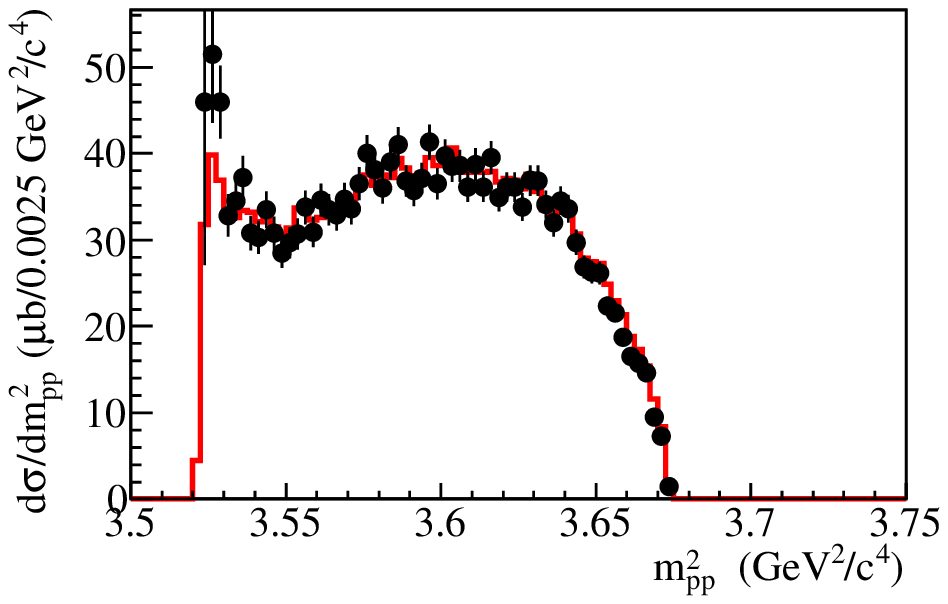}
\includegraphics[clip,width=6.2cm]{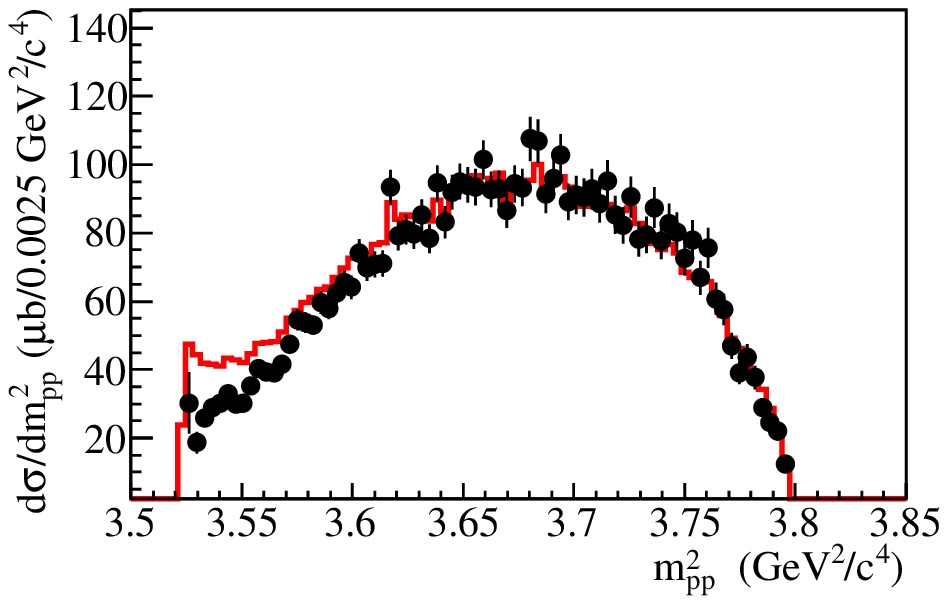}
\includegraphics[clip,width=6.2cm]{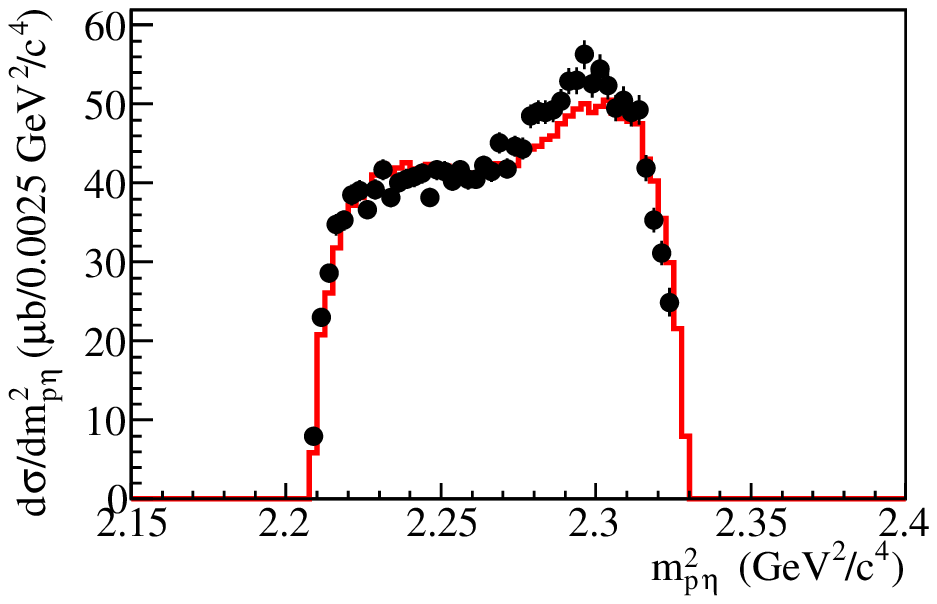}
\includegraphics[clip,width=6.2cm]{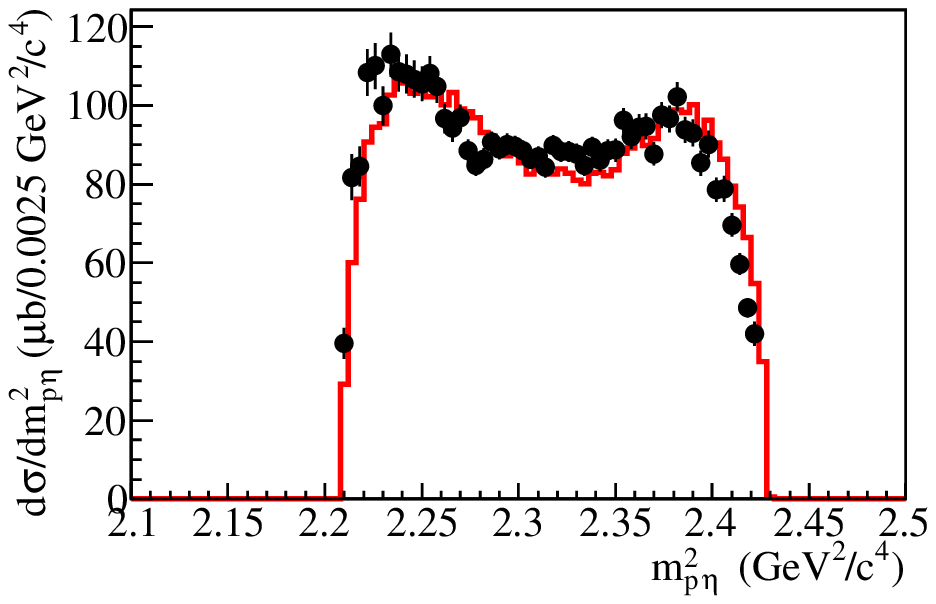}
\caption{Invariant mass distributions for the $pp\to pp\eta$ reaction
at (left) $Q=40$~MeV and (right) $Q=72$~MeV~\cite{PET2009} compared
to the best fit on the basis of Eq.~(\ref{msq}). \label{masses}}
\end{center}
\end{figure}

\begin{figure}[htb]
\begin{center}
\includegraphics[clip,width=6.2cm]{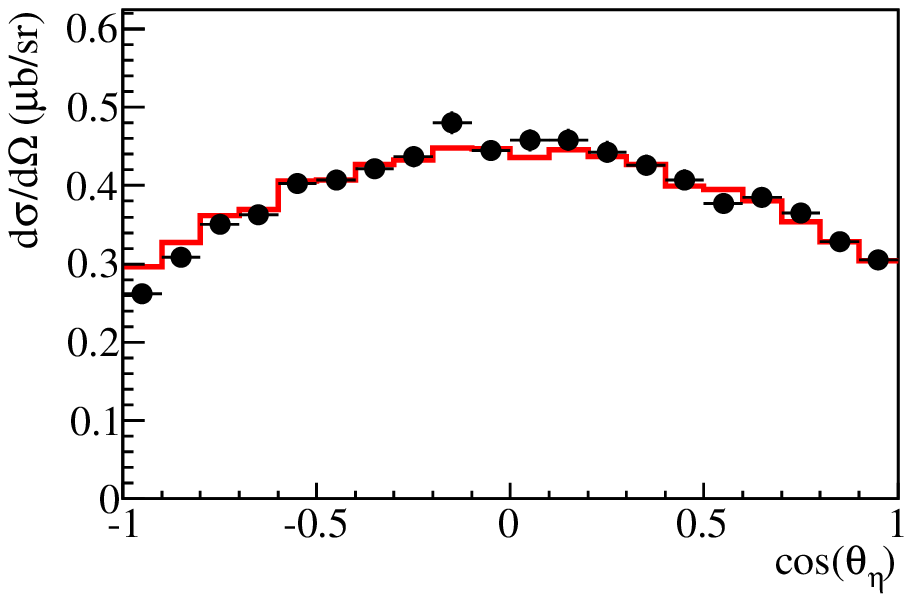}
\includegraphics[clip,width=6.2cm]{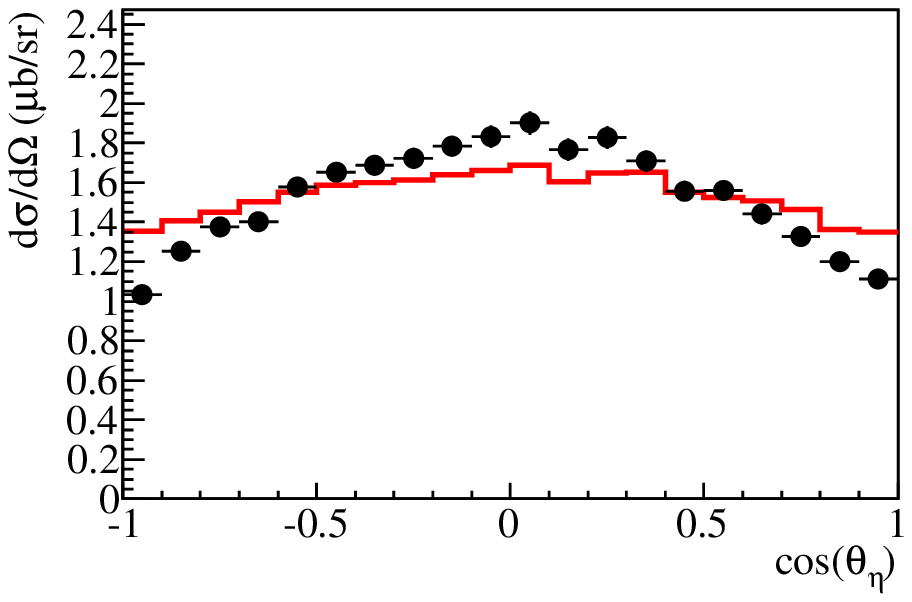}
\caption{Angular distributions of the $\eta$ from the $pp\to pp\eta$
reaction at (left) $Q=40$~MeV and (right) $Q=72$~MeV~\cite{PET2009}
compared to the best fit on the basis of Eq.~(\ref{msq}).
\label{angles}}
\end{center}
\end{figure}

The invariant mass distributions shown in Fig.~\ref{masses} are well
reproduced at 40~MeV but for the 72~MeV data, where the statistics
are higher, the description in the \textit{FSI} region is poor. This
better description of the data at 40~MeV persists also for the $\eta$
angular distributions shown in Fig.~\ref{angles}. These are clearly
more bowed than the earlier COSY-TOF results at
41~MeV~\cite{ABD2003}. In the simple model of Eq.~(\ref{msq}),
deviations from isotropy must come from $Ss$-$Sd$ interference and
these must be too small in the model at 72~MeV.

\begin{figure}[htb]
\begin{center}
\includegraphics[clip,width=12.4cm]{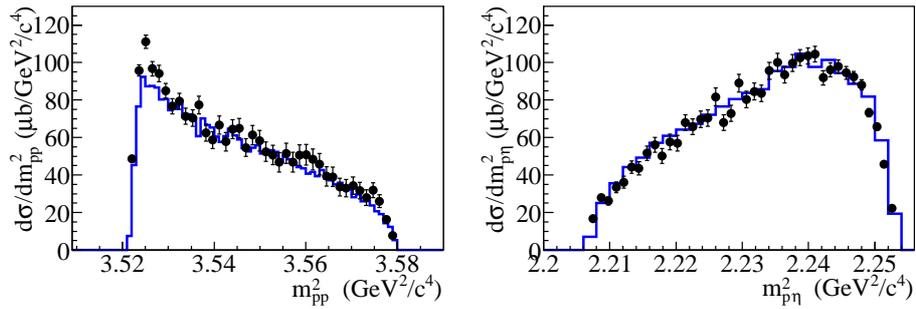}
\caption{Invariant mass distributions for the $pp\to pp\eta$ reaction
at $Q=15$~MeV~\cite{MOS2004} compared to the predictions of
Eq.~(\ref{msq}), using the standard parameters. \label{COSY11}}
\end{center}
\end{figure}

If higher partial waves cause some problems at large excess energies,
what happens at lower values of $Q$? In fact, with the parameters
tuned to describe the 40 and 72~MeV data, the shapes of the COSY-11
data at 15.5 and 10~MeV are well described, as can be judged from the
results shown in Fig.~\ref{COSY11}. In this analysis, higher partial
waves in the $pp$ system are vital for the description of the data at
the largest $s_{pp}$; it the significant contribution from the $Ps$
wave that gives more events at high $m_{pp}$ and hence low
$m_{p\eta}$. Since there is no associated angular dependence, this is
NOT a proof and an unambiguous separation of $Ps$ from $Ss$ would
require a measurement of the initial spin-spin correlation parameter.

\begin{figure}[htb]
\begin{center}
\includegraphics[clip,width=9cm]{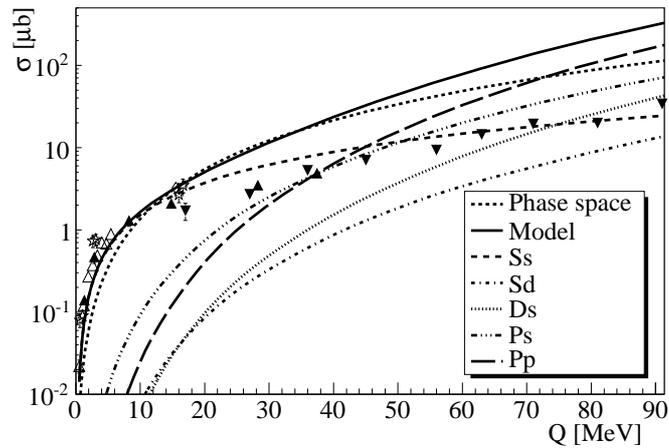}
\caption{Total cross sections for the $pp\to pp\eta$ reaction
compared to the predictions of Eq.~(\ref{msq}), using the standard
parameters. The contributions from individual partial waves within
the model are indicated. The experimental data are taken from
Refs.~\cite{CAL1997,MOS2004,CAL1996,SMY2000,HIB1998,CHI1994}.\label{stot}}
\end{center}
\end{figure}

If the only energy dependence of the partial wave amplitudes were
through the kinematic and \textit{FSI} factors of Eq.~(\ref{msq}),
the energy dependence of the total cross section could be predicted.
This is shown in Fig.~\ref{stot} along with the contributions from
individual partial waves. Since the scaling is arbitrary, we do not
know if disagreement is due to a \textit{FSI} effect at small excess
energy or the constancy ansatz at large $Q$. In the $pn\to d\eta$
case the \textit{FSI} effect extends up to 10~MeV and this could
reach even further for $pp \to pp\eta$ because the $pp$ subsystem
could then take up some of the energy.

%
%
\section{The quasi-two-body $\boldsymbol{pp \to pp\eta(\eta^{\prime})}$ reactions}

To study the $\eta pp$ \textit{FSI} experimentally, we need to
control to some extent the higher partial waves. The $\eta d$ case is
simpler because it is a two-body system, which leads me to my final
points.

\begin{figure}[htb]
\begin{center}
\includegraphics[clip,angle=0,width=7cm]{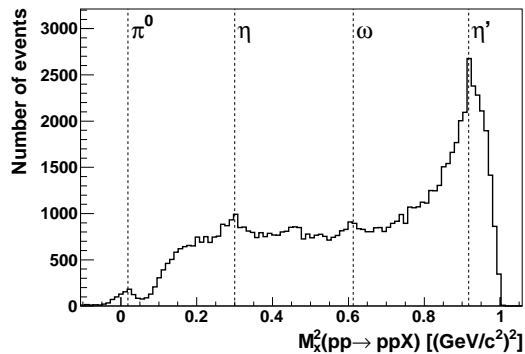}
\caption{Missing-mass spectrum in the $pp\to ppX$ reaction at
2.57~GeV from preliminary ANKE data, where a cut of 3~MeV has been
placed on the excitation energy in the recoiling $pp$
system.\label{etaprime}}
\end{center}
\end{figure}

The ANKE spectrometer has only limited acceptance, but it can measure
well $pp \to \{pp\}_SX$, where the diproton $\{pp\}_S$ has an
excitation energy below 3~MeV, so that it is dominantly in the
$^{1\!}S_0$ state. Data have already been published for the $\eta$ at
$Q\approx 55$~MeV~\cite{DYM2009} and it is seen from the preliminary
missing-mass data at 2.57~GeV shown in Fig.~\ref{etaprime} that there
is a clear $\eta^{\prime}$ signal, also at $Q\approx 55$~MeV.
Approximately 500 events are expected to be extracted from the
signal.

\begin{figure}[htb]
\begin{center}
\includegraphics[clip,width=7cm]{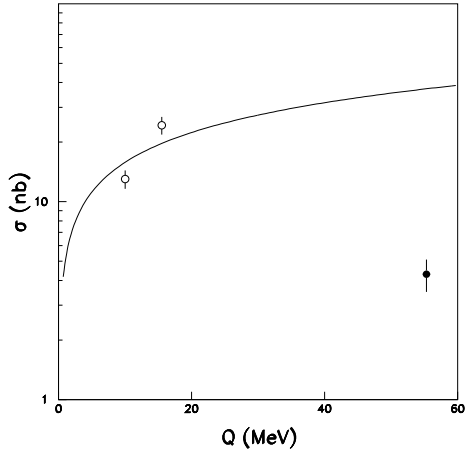}
\caption{Cross section for $pp \to \{pp\}_S\eta$ in the ANKE
conditions where $E_{pp}<3$~MeV and $\cos\theta_{\eta}>0.95$. The
open circles are from COSY-11 data~\cite{MOS2004,KLA2010} and the
closed one from ANKE~\cite{DYM2009}. The curve represents a simple
phase-space dependence $\propto\sqrt{Q}$. \label{sigma_etaprime}}
\end{center}
\end{figure}

There are, as yet, very few measurements in the domain of the ANKE
kinematics, where $E_{pp}<3$~MeV and $\cos\theta_{\eta}>0.95$.
However, points at $Q=15.5$~\cite{MOS2004} and 10~MeV~\cite{KLA2010}
could also be extracted from COSY-11 data and these are shown in
Fig.~\ref{sigma_etaprime} along with a phase-space $\sqrt{Q}$
dependence. The deviations from $\sqrt{Q}$ at small $Q$ are due, in
part, to the $\{pp\}_S$ system not having a fixed mass, but there
must be Physics in the behaviour between 15.5 and 55~MeV. More data
of this kind in the low $E_{pp}$ region are necessary to establish
the extent of the $\eta\{pp\}_S$ \textit{FSI}.

\section{Conclusions}

Three new data sets have here been presented on $\eta$ production in
nucleon-nucleon collisions. Preliminary ANKE results seem to show
some $\eta d$ \textit{FSI}, thought perhaps not as strong as CELSIUS
and much weaker than that for $\eta^3$He. On the other hand, one must
be cautious about the assumption that these sub-threshold data can be
interpreted purely in terms of the single-scattering approximation.

The broad conclusions that might be drawn from the CELSIUS $pp\to
pp\eta$ data are that the valleys in the Dalitz plots demonstrate
that higher partial waves, at least $Pp$, are necessary at both 40
and 72~MeV. The model with constant amplitudes (apart from kinematic
factors) fitted to the 40~MeV data describes very well the 10 and
15.5~MeV results. However, even here one needs more than just
$Ss$-wave and by 72~MeV far higher partial waves may be required. On
the other hand, the proposed parameterisation is far from unique and,
for example, the $Sd$ amplitude has been neglected for an incident
$pp$ $F$-wave. Constant amplitudes, apart from the \textit{FSI} and
kinematic factors, must overestimate the cross section at large $Q$.
The excess in the high mass part of the $s_{pp}$ spectrum is
``explained'' as being due to higher partial waves in the $pp$
system. It should though be noted that a similar shape is seen in the
COSY-11 $pp\to pp\eta^{\prime}$ data~\cite{KLA2010}. Does this mean
that $Ps$ waves enter at the same relative rate for $\eta$ and
$\eta^{\prime}$ production?

In this symposium we are, of course, mainly interested in the
\textit{FSI} of the $\eta$ with nucleons and nuclei and this would
undoubtedly be simpler to study in the $pp$ case if one looked at the
quasi-two-body $pp \to \{pp\}_S\eta$ reaction. The numbers of partial
waves and kinematic variables are vastly reduced and one could then
concentrate on the critical $\eta\{pp\}_S$ relative momentum. For
this we need more data!\\

The material presented here has been derived principally through
collaborations with Henrik Petr\'en and Sergey Dymov. I am also
grateful to Pawel Moskal and Pawel Klaja for providing me with
COSY-11 results. I should like to thank the organisers for meeting
support.

%
%

\end{document}